\documentclass[aps,showpacs,preprint]{revtex4}

\usepackage{amsmath}  
\usepackage{amssymb}  
\usepackage{graphicx} 

\begin{document}

\title{Heterojunction organic photovoltaic cells as molecular heat engines: A simple model for the performance analysis}
\author{Mario Einax$^{1,2}$}
\email{meinax@uos.de}
\author{Marcel Dierl$^{1}$}
\author{Abraham Nitzan$^{2}$}
\affiliation{ $^{1}$Fachbereich Physik, Universit\"at Osnabr\"uck,
Barbarastra{\ss}e 7, 49076 Osnabr\"uck, Germany\\
$^{2}$School of Chemistry, Tel Aviv University, Tel
Aviv 69978, Israel}



\begin{abstract}
  Organic heterojunction solar cells are analyzed within a minimal
  model that includes the essential physical features of such
  systems. The dynamical properties of this model, calculated using a
  master equation approach, account for the qualitative behavior of
  such systems. The model yields explicit results for current-voltage
  behavior as well as performance characteristics expressed in
  terms of the thermodynamic efficiency as well as the power conversion
  efficiency at maximum power, making it possible to evaluate the
  optimal setup for this device model.
\end{abstract}

\maketitle

\section{Introduction}
The limited supply of today's main energy sources, such as oil and
coal, will force us to rely increasingly on renewable energy sources.
Besides wind and water power, energy conversion based on photovoltaic (PV) devices has
received much attention as such a source. Promising systems for next
generation devices are organic photovoltaic solar cells (OPV) because of their potential
for low-cost processing. Of particular interest are polymer-based
heterojunctions consisting of a blend of electron-donor (D) and
electron-acceptor (A) material (for reviews, see
Refs.~\cite{Deibel/Dyakonov:2010,Nicholson/Castro:2010}, and references
therein). A prominent acceptor material is the buckminsterfullerene
(C$_{60}$)
\cite{Saricifti/etal:1992,Yu/etal:1995,Hoppe/etal:2004,Koeppe/etal:2006}.
The quest to improve the energy conversion efficiency of such systems
is the focus of intensive current research.

To evaluate and subsequently improve the efficiency of organic
photovoltaic cells it is crucial to understand the underlying energy
conversion processes and how material properties affect their overall
performance. Widely accepted is the multistep generation process that
starts with photon absorption by the donor (often a polymer) yielding
an exciton (bounded electron-hole pair). The generated exciton
diffuses to the D-A interface, where it dissociates into free charge
carriers, which are later transported to the electrodes. The D-A
interface should be constructed to favor energetically fast and
efficient electron transfer leading to exciton dissociation.

The dynamics of electron transfer at the D-A interface is of crucial
importance for the performance of heterojunction solar cells as
measured by their efficiency. In considering this issue, one may
address the thermodynamic efficiency $\eta^*$ \cite{Rutten/etal:2009,Giebink/etal:2011}
by considering the solar cell as a heat engine operating between a hot
and a cold reservoir with temperatures $T_{\rm\scriptscriptstyle S}$ (``sun temperature'',
representing the incident radiation) and $T$ (temperature of the chemical environment),
respectively. Alternatively, the conversion efficiency $\eta$
\cite{Potscavage/etal:2009} associated with the maximum power point in
the current-voltage \mbox{($J$\,-\,$U$)} characteristic is of interest as a realistic
performance measure. Establishing the
relationship between system properties that affect exciton
dissociation at the D-A interface and the cell efficiency is a major
goal of the ongoing research. Within this effort, it is useful to
consider simple model systems for which one can obtain explicit
relationships between system structure and characteristic parameters and its
performance measures. In this article, we describe and analyze such a
model system.

Our model consists of coupled donor and acceptor molecules, each
described as a two level (highest occupied molecular level, HOMO, and
lowest unoccupied molecular level, LUMO) system, situated between two
electrodes. As such, it is an extension of a simpler model recently
analyzed in a similar context by Rutten et al. \cite{Rutten/etal:2009}
but with an important additional feature - the existence of an heterojunction
characterized by energetic parameters - Coulomb interaction and
donor-acceptor LUMO-LUMO gap, that were identified as important
driving factors in the operation of such systems. The system dynamics
associated with this model is described by a kinetics scheme derived
using a lattice gas
approach, \cite{Einax/etal:2010a,Einax/etal:2010b,Dierl/etal:2011}
similar in spirit to previous work
\cite{Sylvester-Hivid/etal:2004,Nelson/etal:2004,Burlakov/etal:2005,Lei/etal:2008,Wagenpfahl/etal:2010}
that uses a master equation approach to analyze cell dynamics.

We show that an effective mechanism for both exciton pair formation
and dissociation can be captured by a minimal model of this type,
which can by used as a framework for discussing the current- and
power-voltage curves and the cell efficiency. In particular, the model
leads directly to the predictions of an optimal interface energy gap
$\Delta\varepsilon$ (usually associated with the energy difference
between the lowest unoccupied molecular levels (LUMO's) of the D and A
molecules) for these efficiency measures. This
can be compared with the performance of an ideal device, in which nonradiative losses are absent. In
this ideal case, the thermodynamic efficiency is found to decrease monotonously, whereas the power
conversion efficiency still goes through a maximum, with increasing
$\Delta\varepsilon$.

\begin{figure}[t!]
\centering
 \includegraphics[width=0.5\textwidth]{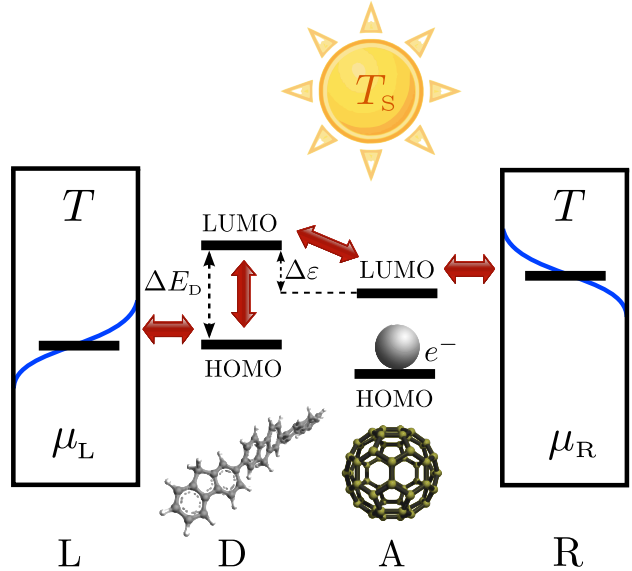}
 \caption{A minimal model of an organic heterojunction PV cell. The
   system consists of one donor (e.g. a suitable polymer) and one
   acceptor, e.g. fullerene sites, each characterized by their HOMO
   and LUMO levels.}
 \label{fig:fig1}
\end{figure}

\begin{figure}[t!]
\centering
 \includegraphics[width=0.35\textwidth]{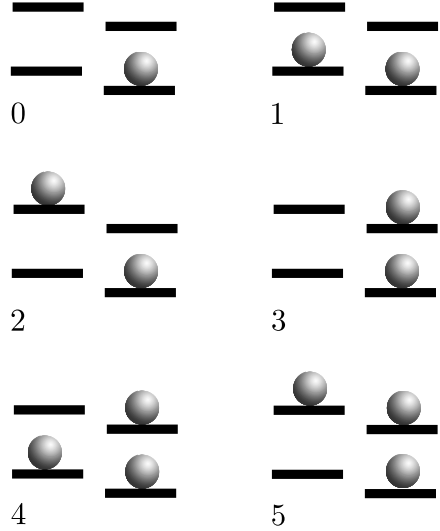}
 \caption{The six accessible microstates of the model system
   considered. The probability to find the system in state $j$\, ($j =
   0,...,5$) is denoted $P_j$.}
 \label{fig:fig2}
\end{figure}

\section{Model and Computational Details}
\subsection{Model}
The PV cell model considered here is a one-dimensional lattice gas with two
``effective'' sites $l=\rm D,\,A$. One site represents the donor ($\rm
D$, e.~g., polymer-based material) and the second acts as acceptor
($\rm A$, e.~g., fullerene-based material). Each of the sites is
represented as a two-state system with energy levels
($\varepsilon_{\rm\scriptscriptstyle
  D1},\varepsilon_{\rm\scriptscriptstyle D2}$) and
($\varepsilon_{\rm\scriptscriptstyle
  A1},\varepsilon_{\rm\scriptscriptstyle A2}$) corresponding to the
(HOMO, LUMO) levels of the donor and acceptor species, respectively. In
what follows we use the notation $\Delta E_l = \varepsilon_{l2} -
\varepsilon_{l1}$ ($l=\rm D,\,A$), for the energy differences that represent the donor and
acceptor band gaps, and refer to
$\Delta\varepsilon = \varepsilon_{\rm\scriptscriptstyle D2} -
\varepsilon_{\rm\scriptscriptstyle A2}$ as the interface or
donor-acceptor LUMO-LUMO gap. The microstates of the system can be
specified by the set of occupation numbers $\boldsymbol{n} =
(n_{\rm\scriptscriptstyle D1},n_{\rm\scriptscriptstyle
  D2},n_{\rm\scriptscriptstyle A1},n_{\rm\scriptscriptstyle A2})$,
where $n_{\rm\scriptscriptstyle lj}= 0$ or $1$ ($l=D,A$; $j=1,2$)
if the corresponding level is vacant or occupied by an electron.

To assign further realistic contents to this model, we introduce the
restrictions $n_{\rm\scriptscriptstyle D1} n_{\rm\scriptscriptstyle
  D2} = 0$ and $n_{\rm\scriptscriptstyle A1} =1$. The first of these
restrictions implies that the donor can be in the ground or excited
state, or, following charge separation, in positively ionized state,
but excludes its doubly occupied (negatively charged) state. The
excited donor state $(n_{\rm\scriptscriptstyle
  D1},\,n_{\rm\scriptscriptstyle D2})=(0,\,1)$ represents the exciton
formed as result of light absorption. The second condition implies
that the acceptor can be in either its ground
$(n_{\rm\scriptscriptstyle A1},\,n_{\rm\scriptscriptstyle
  A2})=(1,\,0)$ or its negative ion $(n_{\rm\scriptscriptstyle
  A1},n_{\rm\scriptscriptstyle A2})=(1,\,1)$ states. Therefore, the system
is characterized by six states with respect to the occupations
$(n_{\rm\scriptscriptstyle D1},\,n_{\rm\scriptscriptstyle
  D2},\,n_{\rm\scriptscriptstyle A1},\,n_{\rm\scriptscriptstyle A2})$,
that we denote by the integers $0,\,...,\,5$ as shown in
Table~\ref{table:state}.

\begin{table}
  \caption{System states and their occupations with corresponding energies.}
  \label{table:state}
  \begin{tabular}{|l|l|l|}
    \hline
    STATE & OCCUPATION & ENERGY \\[0ex]
    & $(n_{\rm\scriptscriptstyle D1},\,n_{\rm\scriptscriptstyle D2},\,n_{\rm\scriptscriptstyle A1},\,n_{\rm\scriptscriptstyle A2})$ & \\
    \hline
    $0$ & $(0,0,1,0)$ & $\varepsilon_0 = \varepsilon_{\rm\scriptscriptstyle A1}$ \\
    \hline
    $1$ & $(1,0,1,0)$ & $\varepsilon_1 = \varepsilon_{\rm\scriptscriptstyle D1} + \varepsilon_{\rm\scriptscriptstyle A1}$\\
    \hline
    $2$ & $(0,1,1,0)$ & $\varepsilon_2 = \varepsilon_{\rm\scriptscriptstyle D2} + \varepsilon_{\rm\scriptscriptstyle A1}$\\
    \hline
    $3$ & $(0,0,1,1)$ & $\varepsilon_3 = \varepsilon_{\rm\scriptscriptstyle A1} + \varepsilon_{\rm\scriptscriptstyle A2}+V_{\rm\scriptscriptstyle C} + V_{\rm\scriptscriptstyle C}'$ \\
    & & $\hspace*{0.45cm} =\varepsilon_{\rm\scriptscriptstyle A1} + \tilde\varepsilon_{\rm\scriptscriptstyle A2}+V_{\rm\scriptscriptstyle C}'$ \\
    \hline
    $4$ & $(1,0,1,1)$ & $\varepsilon_4 = \varepsilon_{\rm\scriptscriptstyle D1} + \varepsilon_{\rm\scriptscriptstyle A1} + \varepsilon_{\rm\scriptscriptstyle A2} +V_{\rm\scriptscriptstyle C}$ \\
    & & $\hspace*{0.45cm} =\varepsilon_{\rm\scriptscriptstyle D1} + \varepsilon_{\rm\scriptscriptstyle A1} + \tilde\varepsilon_{\rm\scriptscriptstyle A2}$\\
    \hline
    $5$ & $(0,1,1,1)$ & $\varepsilon_5 = \varepsilon_{\rm\scriptscriptstyle D2} + \varepsilon_{\rm\scriptscriptstyle A1} + \varepsilon_{\rm\scriptscriptstyle A2}+V_{\rm\scriptscriptstyle C}$\\
    & & $\hspace*{0.45cm} =\varepsilon_{\rm\scriptscriptstyle D2} + \varepsilon_{\rm\scriptscriptstyle A1} + \tilde\varepsilon_{\rm\scriptscriptstyle A2}$ \\
    \hline
  \end{tabular}
\end{table}
In the expressions for the states energies, $V_{\rm\scriptscriptstyle
  C} > 0$ is the Coulombic repulsion between two electrons on the
acceptor, whereas $V_{\rm\scriptscriptstyle C}' > 0$ is the Coulombic
energy cost to move an electron away from the hole remaining on the
donor. In general, we expect that $V_{\rm\scriptscriptstyle C} >
V_{\rm\scriptscriptstyle C}'$. The sum $V_{\rm\scriptscriptstyle C} +
V_{\rm\scriptscriptstyle C}'$ can be thought of as the exciton binding
energy in this model: It is the total Coulomb energy cost for
dissociating the exciton on the donor by moving an electron to the
acceptor. It is convenient to redefine
$\tilde\varepsilon_{\rm\scriptscriptstyle A2}=
\varepsilon_{\rm\scriptscriptstyle A2}+V_{\rm\scriptscriptstyle C}$ so
that the energies $\varepsilon_{j}$\,($j=0,...,5$) are determined by
the five energy parameters $\varepsilon_{\rm\scriptscriptstyle D1}$,
$\varepsilon_{\rm\scriptscriptstyle D2}$,
$\varepsilon_{\rm\scriptscriptstyle A1}$,
$\tilde\varepsilon_{\rm\scriptscriptstyle A2}$, and
$V_{\rm\scriptscriptstyle C}'$.

At the left (donor, say) and right (acceptor) ends of the system, the
device is connected with two electrodes represented by free electron
reservoirs at chemical potentials $\mu_{\scriptscriptstyle K}$ ($K
= \rm L,\, R$). This is a highly simplified picture that disregards
electron transport within the donor and acceptor phases. We have opted
to make this simplification to focus on the important step of
interfacial exciton dissociation, but future more realistic treatments
should take these components of the overall dynamics into consideration.
As sketched in Fig.~\ref{fig:fig1}, we use the common picture by which
the left reservoir is assumed to exchange electrons only with the HOMO
level of the donor, whereas the right lead exchange electrons with the
upper level of the acceptor. Direct electronic interaction between system
and reservoirs is not explicitly taken into account, but it is implicit
both in the states of molecular species adjacent to metal electrodes
and in the kinetic charge transfer rates.

Next, we construct the kinetic scheme for the time evolution of the
average occupations $P_j$\, ($j=0,..,5$) associated with this level
structure (see Fig.~\ref{fig:fig2}). In writing these equations, we make the
simplifying assumption that electron exchange between molecules and
metals involve only metal electrons at the electrochemical potentials
$\mu_{\rm\scriptscriptstyle L}$ and $\mu_{\rm\scriptscriptstyle R}$ of
the left and right leads (corresponding to a bias potential
$U=(\mu_{\rm\scriptscriptstyle R}-\mu_{\rm\scriptscriptstyle L})/|e|$
where $e$ is the electron charge. We also disregard possible
environmental relaxation dynamics due to polarization effects
associated with the formation of transient charged molecular species.
Generalizing this dynamical picture to take such processes into account
(e.g. by considering time dependent site energies \cite{Arkhipov/etal:1999})
is another important subject for future work. The kinetics process in our
scheme then corresponds to the following processes:
\begin{itemize}
\item[(a)] Electron transfer between levels $\rm D1$ and the left
  electrode and between $\rm A2$ and the right electrode, with rates
  determined by the corresponding molecular energies, molecules-leads coupling,
  electrochemical potentials in the leads, and the environmental temperature, $T$.
\item[(b)] Electron transfer between donor and acceptor, determined by the coupling between them,
  the corresponding state energies, and the temperature, $T$.
\item[(c)] Light induced electron excitation (rate $k_s$) and
  relaxation (rate $\tilde{k}_{s}$) between donor levels $\rm D1$ and
  $\rm D2$. These rates are modeled as thermal rates determined by
  the corresponding state energies and radiative coupling, and
  the ``sun temperature'' $T_{s}$.
\item[(d)] Radiationless (thermal) electron transitions between level
  $\rm D1$ and $\rm D2$ with rates $k_{\rm nr}$ (excitation) and
  $\tilde k_{\rm nr}$ (relaxation), determined as in (c) except that relevant
  coupling is vibronic in origin and the temperature involved is the
  environmental temperature $T$.
\end{itemize}
Explicitly, the transition rates $k_{j'\,j} = k_{j' \leftarrow j}$
from state $j$ to $j'$ are given by
\begin{align}
  \label{eq:kL1}
    k_{01}&= k_{34} \equiv \tilde k_{\rm\scriptscriptstyle L} = \nu_{\rm \scriptscriptstyle{L}} \tilde f(x_{\rm \scriptscriptstyle{L}})\, ,\\
\label{eq:kL2}
    k_{10}&= k_{43} \equiv k_{\rm\scriptscriptstyle L} = \nu_{\rm \scriptscriptstyle{L}}  f(x_{\rm \scriptscriptstyle{L}} )\, ,\\
\label{eq:ksnr1}
    k_{12}&= k_{45} \equiv \tilde k_{\rm\scriptscriptstyle S} + \tilde k_{\rm nr}\nonumber\\
    &= \nu_{\rm \scriptscriptstyle{S}} (1 + n_{\rm \scriptscriptstyle{S}} (x_{\rm \scriptscriptstyle{S}}))
    + \nu_{\rm nr} (1 + n_{\rm nr} (x_{\rm nr}))\, ,\\
\label{eq:ksnr2}
    k_{21}&= k_{54} \equiv k_{\rm\scriptscriptstyle S} + k_{\rm nr} = \nu_{\rm \scriptscriptstyle{S}} n_{\rm \scriptscriptstyle{S}} (x_{\rm \scriptscriptstyle{S}}) + \nu_{\rm nr} n_{\rm nr} (x_{\rm nr})\, ,\\
\label{eq:kpn1}
    k_{32}&\equiv \tilde k_{\rm\scriptscriptstyle DA} = \nu_{\rm\scriptscriptstyle DA} \tilde f(x_{\rm\scriptscriptstyle DA})\, ,\\
\label{eq:kpn2}
    k_{23}&\equiv k_{\rm\scriptscriptstyle DA} = \nu_{\rm\scriptscriptstyle DA} f(x_{\rm\scriptscriptstyle DA})\, ,\\
\label{eq:kR1}
    k_{03}&= k_{14}= k_{25} \equiv \tilde k_{\rm\scriptscriptstyle R} = \nu_{\rm \scriptscriptstyle{R}}  \tilde f(x_{\rm \scriptscriptstyle{R}})\, ,\\
\label{eq:kR2}
    k_{30}&= k_{41}= k_{52} \equiv k_{\rm\scriptscriptstyle R} = \nu_{\rm \scriptscriptstyle{R}}  f(x_{\rm \scriptscriptstyle{R}})\, .
\end{align}
In these equations $f(x) = 1/\left[\exp(x)+1\right]$ and $\tilde
f(x)\equiv 1 - f(x)$, $x_{\rm\scriptscriptstyle L} =
(\varepsilon_{\rm\scriptscriptstyle D1} -\mu_{\rm\scriptscriptstyle
  L})/k_{\rm\scriptscriptstyle B}T$, $x_{\rm\scriptscriptstyle R} =
(\tilde\varepsilon_{\rm\scriptscriptstyle A2}
-\mu_{\rm\scriptscriptstyle R})/k_{\rm\scriptscriptstyle B}T$, and
$x_{\rm\scriptscriptstyle DA} = (\varepsilon_{\rm\scriptscriptstyle
  D2} -\tilde\varepsilon_{\rm\scriptscriptstyle A2} -
V_{\rm\scriptscriptstyle C}')/k_{\rm\scriptscriptstyle B}T$. The rate coefficients $\nu$
are determined by the corresponding couplings. Following Ref.~\cite{Rutten/etal:2009}, the transition rates between the HOMO and
LUMO in the donor phase are assigned by using the boson population
factors $n_{\rm{\scriptscriptstyle S}} = 1/\left[\exp(x_{\rm
    {\scriptscriptstyle S}})-1\right]$ and $n_{\rm{nr}} =
1/\left[\exp(x_{\rm nr})-1\right]$ with scaled energies
$x_{\rm\scriptscriptstyle S} = \Delta E_{\rm\scriptscriptstyle
    D}/k_{\rm\scriptscriptstyle B}T_{\rm\scriptscriptstyle S}$ and
$x_{\rm nr} = \Delta E_{\rm\scriptscriptstyle
    D}/k_{\rm\scriptscriptstyle B}T$. $\tilde
k_{\rm\scriptscriptstyle L}\equiv k_{01} = k_{34}$ is the rate to move
an electron from level $\rm D1$ to the left electrode, $\tilde
k_{\rm\scriptscriptstyle R}\equiv k_{03} = k_{14} = k_{25}$ is
similarly the rate from $\rm A2$ to the right electrode.
$k_{\rm\scriptscriptstyle L}\equiv k_{10} = k_{43}$ and
$k_{\rm\scriptscriptstyle R}\equiv k_{30} = k_{41} = k_{52}$ are the
corresponding reverse rates. The transition rate from $\rm D2$ to $\rm A2$
is given by $\tilde k_{\rm\scriptscriptstyle DA}\equiv k_{32}$, whereas the opposite rate is
$k_{\rm\scriptscriptstyle DA}\equiv k_{23}$. Also, we denote by
$k_{\rm\scriptscriptstyle S}$ and $k_{\rm nr}$ the radiative and thermal
excitation rates in the donor and by $\tilde k_{\rm\scriptscriptstyle
  S}$ and $\tilde k_{\rm nr}$ the corresponding relaxation rates, so
$k_{\rm\scriptscriptstyle S} + k_{\rm nr} \equiv k_{21} = k_{54}$ and
$\tilde k_{\rm\scriptscriptstyle S} + \tilde k_{\rm nr} \equiv k_{12}
= k_{45}$.

\subsection{Computational Details}
The master equation with the rates Eqs.~(1)-(8) that describe
the time evolution of the probabilities
$P_j = P(n_{\rm\scriptscriptstyle D1},\,n_{\rm\scriptscriptstyle
  D2},\,n_{\rm\scriptscriptstyle A1},\,n_{\rm\scriptscriptstyle A2})$
($j=0,\,...,\,5$) to be in the six possible states thus reads
\begin{align}
\label{eq:dP_0}
 \frac{dP_{0}(t)}{dt}&=k_{01} P_{1}(t) + k_{03} P_{3}(t) -
  \left( k_{10}+k_{30}\right)P_{0}(t)\, ,\\
\label{eq:dP_1}
  \frac{dP_{1}(t)}{dt}&=k_{10} P_{0}(t) + k_{14} P_{4}(t) + k_{12} P_{2}(t) \nonumber \\
  &\hspace{3 ex}- \left( k_{01}+k_{41}+k_{21} \right) P_{1}(t)\, ,\\
\label{eq:dP_2}
  \frac{dP_{2}(t)}{dt}&=k_{21} P_{1}(t) + k_{23} P_{3}(t) + k_{25} P_{5}(t) \nonumber \\
  &\hspace{3 ex}- \left( k_{12}+k_{32} + k_{52} \right) P_{2}(t)\, ,\\
\label{eq:dP_3}
  \frac{dP_{3}(t)}{dt}&=k_{30} P_{0}(t) + k_{32} P_{2}(t) + k_{34} P_{4}(t) \nonumber \\
  &\hspace{3 ex}- \left( k_{03}+k_{23}+k_{43}\right) P_{3}(t)\, , \\
\label{eq:dP_4}
  \frac{dP_{4}(t)}{dt}&=k_{43} P_{3}(t) + k_{41} P_{1}(t) + k_{45} P_{5}(t) \nonumber \\
  &\hspace{3 ex}- \left( k_{34}+k_{14}+k_{54}\right) P_{4}(t)\, ,
\end{align}
and normalization implies that
\begin{equation}
\label{eq:norm}
 P_{5}(t)=1-\sum\limits_{j=0}^4 P_j(t)\, .
\end{equation}
In terms of these probabilities electron currents can be expressed as
\begin{align}
\label{eq:current1}
   J_{\rm\scriptscriptstyle L}(t) &=  k_{\rm\scriptscriptstyle L} ( P_{0} + P_{3})
         - \tilde k_{\rm\scriptscriptstyle L} ( P_{1} + P_{4}) \\
\label{eq:current2}
   J_{\rm\scriptscriptstyle R}(t) &= \tilde k_{\rm\scriptscriptstyle R} ( P_{3} + P_{4} + P_{5} ) - k_{\rm\scriptscriptstyle R} ( P_{0} + P_{1} + P_{2} ) \\
\label{eq:current3}
   J_{\rm\scriptscriptstyle S}(t) &= k_{\rm\scriptscriptstyle S} ( P_{1} + P_{4}) - \tilde k_{\rm\scriptscriptstyle S} ( P_{2} + P_{5}) \\
\label{eq:current4}
   J_{\rm nr}(t) &= k_{\rm nr} (P_{1} + P_{4} ) - \tilde k_{\rm nr} (P_{2} + P_{5}) \\
\label{eq:current5}
   J_{\rm\scriptscriptstyle DA}(t) &= \tilde k_{\rm\scriptscriptstyle DA} P_{2} - k_{\rm\scriptscriptstyle DA} P_{3} \, .
\end{align}
$J_{\rm\scriptscriptstyle L}$ ($J_{\rm\scriptscriptstyle R}$) is the current entering (leaving) the molecular system from (to) the electrodes,
$J_{\rm\scriptscriptstyle S}$ and $J_{\rm nr}$ are, respectively, the light induced and nonradiative transition currents between the HOMO and the LUMO in the donor phase, and $J_{\rm\scriptscriptstyle DA}$ is the average current between the donor and acceptor species. Below we will focus on the steady state magnitude, $J_{\rm\scriptscriptstyle L} =
J_{\rm\scriptscriptstyle R} = J_{\rm\scriptscriptstyle DA} =
J_{\rm\scriptscriptstyle S} + J_{\rm nr} = J$, of these currents.

The time evolution and steady state currents associated with this kinetic model can be
evaluated exactly, however such exact solution becomes costly for larger, more realistic system
that takes into account also transport within the donor and acceptor phases. We therefore advance
also an approximate mean field treatment and show that it can provide a good approximation to the
exact analysis and can be used for realistic multilevel systems at a relatively
low computational cost.

To this end, we introduce the averaged site occupations  $p_{\rm\scriptscriptstyle D1}(t) \equiv \langle n_{\rm\scriptscriptstyle D1} \rangle = P_1(t) +
P_4(t)$, $p_{\rm\scriptscriptstyle D2}(t) \equiv \langle n_{\rm\scriptscriptstyle D2} \rangle = P_2(t) + P_5(t)$, and
$p_{\rm\scriptscriptstyle A2}(t) \equiv \langle n_{\rm\scriptscriptstyle A2} \rangle = P_3(t) + P_4(t) + P_5(t)$ (see also the illustration in Fig.~\ref{fig:fig2}). Therefore, the currents (15)-(19) can be rewritten as follows
\begin{align}
\label{eq:current1_MF}
   J_{\rm\scriptscriptstyle L}(t) &= k_{\rm\scriptscriptstyle L}(\tilde p_{\rm\scriptscriptstyle D1}-p_{\rm\scriptscriptstyle D2}) - \tilde k_{\rm\scriptscriptstyle L} p_{\rm\scriptscriptstyle D1}\, ,\\
\label{eq:current2_MF}
   J_{\rm\scriptscriptstyle R}(t) &= \tilde k_{\rm\scriptscriptstyle R} p_{\rm\scriptscriptstyle A2} - k_{\rm\scriptscriptstyle R} \tilde p_{\rm\scriptscriptstyle A2}\, ,\\
\label{eq:current3_MF}
   J_{\rm\scriptscriptstyle S}(t) &= k_{\rm\scriptscriptstyle S} p_{\rm\scriptscriptstyle D1} - \tilde k_{\rm\scriptscriptstyle S} p_{\rm\scriptscriptstyle D2}\, ,\\
\label{eq:current4_MF}
   J_{\rm nr}(t) &= k_{\rm nr} p_{\rm\scriptscriptstyle D1} - \tilde k_{\rm nr} p_{\rm\scriptscriptstyle D2}\, ,
\end{align}
where $\tilde{p}=1-p$. The treatment of the average current $J_{\rm\scriptscriptstyle DA}(t)$ between the donor and acceptor species is more difficult. In terms of occupation number $J_{\rm\scriptscriptstyle DA}(t)$ it is
\begin{align}
\label{eq:currentDA_OCC}
   J_{\rm\scriptscriptstyle DA}(t) &= \tilde{k}_{\rm\scriptscriptstyle DA} \langle n_{\rm\scriptscriptstyle D2} \left( 1- n_{\rm\scriptscriptstyle A2} \right) \rangle - k_{\rm\scriptscriptstyle DA} \langle \left( 1- n_{\rm\scriptscriptstyle D1} - n_{\rm\scriptscriptstyle D2} \right) n_{\rm\scriptscriptstyle A2} \rangle \, .
\end{align}
Neglecting fluctuations, i.~e., $\langle n_{\rm\scriptscriptstyle D1} n_{\rm\scriptscriptstyle A2}
\rangle \approx \langle n_{\rm\scriptscriptstyle D1} \rangle \langle n_{\rm\scriptscriptstyle A2} \rangle = p_{\rm\scriptscriptstyle  D1}\,p_{\rm\scriptscriptstyle A2}$ and $\langle n_{\rm\scriptscriptstyle D2} n_{\rm\scriptscriptstyle  A2} \rangle \approx \langle n_{\rm\scriptscriptstyle D2} \rangle \langle n_{\rm\scriptscriptstyle A2} \rangle =p_{\rm\scriptscriptstyle D2}\,p_{\rm\scriptscriptstyle A2}$, leads to the mean field expression of the current
\begin{align}
   J^{\rm\scriptscriptstyle MF}_{\rm\scriptscriptstyle DA}(t) &= \tilde k_{\rm\scriptscriptstyle DA} p_{\rm\scriptscriptstyle D2} \tilde p_{\rm\scriptscriptstyle A2} - k_{\rm\scriptscriptstyle DA} (\tilde p_{\rm\scriptscriptstyle D1}-p_{\rm\scriptscriptstyle D2}) p_{\rm\scriptscriptstyle A2} \, .
\label{eq:currentDA_MF}
\end{align}
As a consequence, we arrive at the following mean field rate equations for the averaged site occupations
\begin{align}
\label{eq:rate_equation1}
 \frac{d p_{\rm\scriptscriptstyle D1}}{dt} &= J_{\rm \scriptscriptstyle L}(t) - J_{\rm \scriptscriptstyle S}(t) - J_{\rm nr}(t)\, ,\\
\label{eq:rate_equation2}
 \frac{d p_{\rm\scriptscriptstyle D2}}{dt} &= J_{\rm \scriptscriptstyle S}(t) + J_{\rm nr}(t) - J_{\rm\scriptscriptstyle DA}^{\rm\scriptscriptstyle MF}(t)\, ,\\
\label{eq:rate_equation3}
 \frac{d p_{\rm\scriptscriptstyle A2}}{dt} &= J_{\rm\scriptscriptstyle DA}^{\rm\scriptscriptstyle MF}(t) - J_{\rm\scriptscriptstyle R}(t)\, .
\end{align}
The full kinetics in this approximation is obtained by solving Eqs.~(26)-(28) together with Eqs.~(20)-(23), and (25) self consistently. Note that the number of coupled equations solved in this scheme grows linearly with the number of $N$ of single electron states, while in the exact approach this number is essentially the number of molecular states $\sim 2^N$.

\begin{figure}[t!]
\centering
 \includegraphics[width=0.54\textwidth]{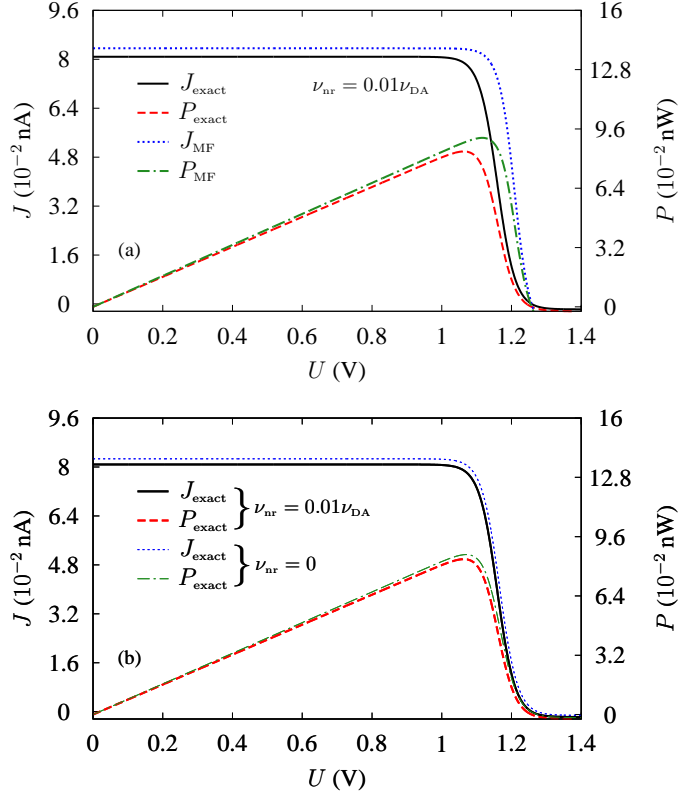}
 \caption{Current ($J$, left vertical axis) and power ($P=UJ$, right vertical axis)
   in the PV cell, plotted against the voltage bias $U$ (see text for parameters).
   Plot (a) compares results from the exact calculation to those obtained from
   mean field approximation. Plot (b) compares the same exact results ($\nu_{\rm nr}=0.01\nu_{\rm\scriptscriptstyle DA}$) to those
   obtained for the ideal cell for which $\nu_{\rm nr}=0$.
   We assume that $\nu_{\rm\scriptscriptstyle DA}=10^{12}$s$^{-1}$.}
 \label{fig:fig3}
\end{figure}

To illustrate the nature of the kinetics that results from these rate processes,
we choose a set of physically reasonable parameters: In the calculation
discussed below the following choice was used:\, $\mu_{\rm\scriptscriptstyle L}=0.0\, \rm eV$,
$\mu_{\rm\scriptscriptstyle R}=\mu_{\rm\scriptscriptstyle L} + |e|U$,
$\varepsilon_{\rm\scriptscriptstyle D1} = -0.1\,\rm eV$,
$\varepsilon_{\rm\scriptscriptstyle D2} = 1.4\,\rm eV$,
$\varepsilon_{\rm\scriptscriptstyle A2} = 0.9\,\rm eV$,
$V_{\rm\scriptscriptstyle C} = 0.25\,\rm eV$, and
$V_{\rm\scriptscriptstyle C'} = 0.15\,\rm eV$. Thus,
$\Delta E_{\rm\scriptscriptstyle D} = \varepsilon_{\rm\scriptscriptstyle D2} -
\varepsilon_{\rm\scriptscriptstyle D1} = 1.5\, \rm eV$ \cite{Soci/etal:2007} and
$V_{\rm\scriptscriptstyle C} + V_{\rm\scriptscriptstyle C'} = 0.4\,\rm eV$
\cite{Gregg/etal:2003,Pensack/etal:2010}. For the temperatures we take $T=300 \rm K$ and $T_s=6000 \rm K$.
The kinetic rates are set to
$\nu_{\rm\scriptscriptstyle L} = \nu_{\rm\scriptscriptstyle R} =
\nu_{\rm\scriptscriptstyle S} = \nu_{\rm nr} = 0.01 \nu_{\rm\scriptscriptstyle DA}$ and $\nu_{\rm\scriptscriptstyle DA}=10^{12}$s$^{-1}$,
describing a system with efficient and fast donor-to-acceptor electron transfer (which can occur on the ps timescale \cite{Rice:1996}),
and moderate radiationless losses, as would be used in such applications.
Finally, note that in the particular example employed here we have considered a situation where the imposed potential bias
falls between the acceptor species and the right electrode. It should be emphasized
that although the results shown in Figs.~3 and 4 are based on these choices,
the qualitative behavior discussed below holds for a wide range of these parameters.

\section{Results and Discussions}
Figure \ref{fig:fig3} shows results for the stationary donor $\rightarrow$ acceptor current obtained from both the
exact solution from Eqs.~(9)-(19) and the mean field approximation, Eqs.~(20)-(28).
Figure \ref{fig:fig3}a compares the results of the mean field approach to their exact counterpart, showing that the former
provides a good approximation to the exact behavior. Figure \ref{fig:fig3}b compares the performance of a junction characterized
by the prescribed parameters to the corresponding ideal junction in which nonradiative losses are absent ($\nu_{\rm nr}=0$).
In both cases, the current is constant until the electrochemical potential on the right electrode comes within
$\sim k_{\rm\scriptscriptstyle B}T$ of the acceptor level A2, and decreases sharply after it exceeds this level.
Consequently, the generated power, $P(U)=U\,J(U)$ goes through a sharp maximum,
$P_{\rm max} = U_{\rm max}\,J(U_{\rm max})$ in that voltage region.

Next, consider the efficiency. The maximal power conversion efficiency is defined by
\begin{align}
\label{eq:Power_Conversion_efficiency}
\eta &= \frac{P_{\rm max}}{P_{\rm\scriptscriptstyle S}} = \frac{U_{\rm max}\,J(U_{\rm max})}{P_{\rm\scriptscriptstyle S}}\, ,
\end{align}
where $P_{\rm\scriptscriptstyle S}$ is the incident radiant power, a constant independent of the process undergone by the system. The thermodynamic efficiency at maximum power is given by
\begin{align}
\label{eq:Td_efficiency}
\eta^* &\equiv \frac{P_{\rm max}}{\dot{Q}_{\rm\scriptscriptstyle S}} = \frac{U_{\rm max}\,J(U_{\rm max})}{\Delta E_{\rm D}\, J_{\rm \scriptscriptstyle S}(U_{\rm max})} = \frac{U_{\rm max}}{\Delta E_{\rm D}}\left(1 + \frac{J_{\rm  nr}(U_{\rm max})}{J_{\rm\scriptscriptstyle S}(U_{\rm max})}\right)\, ,
\end{align}
where $\dot{Q}_{\rm\scriptscriptstyle S}$ is the net energy absorbed per
unit time from the radiation field, that is, heat absorbed from the hot reservoir,
$\dot{Q}_{\rm\scriptscriptstyle S} = \Delta E_{\rm\scriptscriptstyle D} J_{\rm
\scriptscriptstyle S}$. We note in passing that all processes undergone by our system
are accompanied by well defined energy changes: The heat fluxes associated with
the electron exchange processes between system and leads, are, in the present model
$\dot{Q}_{\rm\scriptscriptstyle L}
= (\varepsilon_{\rm\scriptscriptstyle D1}-\mu_{\rm\scriptscriptstyle
  L}) J_{\rm\scriptscriptstyle L}$ and $\dot{Q}_{\rm\scriptscriptstyle R}
= (\mu_{\rm\scriptscriptstyle R}- \varepsilon_{\rm\scriptscriptstyle
  A2}) J_{\rm\scriptscriptstyle R}$, $\dot{Q}_{\rm nr} = \Delta E_{\rm D} J_{\rm nr}$
is the net heat generation per unit time related to nonradiative relaxation processes,
and $\dot{Q}_{\rm\scriptscriptstyle DA} = -\Delta \varepsilon\,
J_{\rm\scriptscriptstyle DA}$ is the heat flux associated with the electron transfer
at the D-A interface. Energy conservation implies that the overall cell power is
the sum of these fluxes, i.~e.,
$P=\dot{Q}_{\rm\scriptscriptstyle L} + \dot{Q}_{\rm\scriptscriptstyle
  R} + \dot{Q}_{\rm\scriptscriptstyle S} + \dot{Q}_{\rm nr} +
\dot{Q}_{\rm\scriptscriptstyle DA} \equiv ( \mu_{\rm\scriptscriptstyle
  R}-\mu_{\rm\scriptscriptstyle L} ) J = U J$.

\begin{figure}[t!]
\centering
 \includegraphics[width=0.54\textwidth]{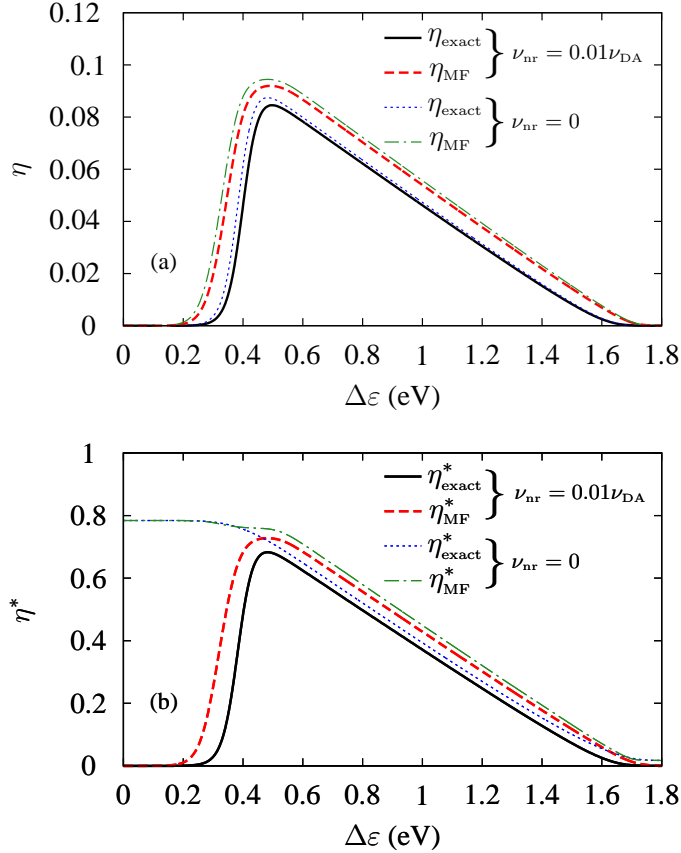}
 \caption{(a) The power conversion efficiency $\eta$ and (b) thermodynamic
   efficiency $\eta^*$ evaluated at maximum power and displayed as functions
   of the interface gap $\Delta\varepsilon$. Mean field results are compared
   with the exact solutions of the underlying master equation.
   Parameters are the same as in Fig.~\ref{fig:fig3} with the difference that
   $\varepsilon_{\rm\scriptscriptstyle A2}$
   runs from $1.4\,\rm eV$ to $-0.4\,\rm eV$ and
   $\mu_{\rm\scriptscriptstyle R}$ is determined by $P_{\rm max}$.
   For the power, $P=P_{\rm\scriptscriptstyle S}$ incident on the cell
   we assume $1.0\,{\rm nW}$ \cite{note}.}
 \label{fig:fig4}
\end{figure}

Of central importance is the dependence of the efficiency on the interface
LUMO-LUMO energy gap $\Delta\varepsilon$. One may intuitively expect to
find that an optimal value of this parameter exists: A finite $\Delta\varepsilon$
is needed to drive the charge separation process but a larger $\Delta\varepsilon$
implies that more energy may be lost to unproductive processes.
Equations (29) and (30) quantify this phenomenon, which is illustrated in Fig.~\ref{fig:fig4}.
Figure \ref{fig:fig4}a shows the power conversion efficiency calculated
for the chosen parameter set using Eq.~(\ref{eq:Power_Conversion_efficiency}).
For comparison, the corresponding result for the ideal cell ($\nu_{\rm nr}=0$)
is also shown. Both are seen to go through a maximum as functions of
the interfacial LUMO-LUMO gap $\Delta\varepsilon$. The thermodynamic efficiency,
Eq.~(\ref{eq:Td_efficiency}), displayed against $\Delta\varepsilon$ in Fig.~\ref{fig:fig4}b,
shows a similar pronounced maximum, however the ideal thermodynamic efficiency is a
monotonously decreasing function of $\Delta\varepsilon$. Note that the ideal cell
efficiency is an upper bound to the actual efficiency, however, for this finite power
operation, it is below the Carnot efficiency,
$\eta_{\rm\scriptscriptstyle C}=1-T/T_{\rm\scriptscriptstyle S} = 0.95$.

Finally, consider again the performance of the mean field approximation relative to the exact solution.
As seen from Figs.~3 and 4, the mean field treatment provides a good approximation that closely
follows the behavior of exact solutions. The influence of correlations is particularly seen in Fig.~\ref{fig:fig4}b,
where we find small but noticeable differences between the mean field and the exact curves for
$\Delta\varepsilon$ beyond the maximum at $\sim 0.4\,\rm eV$.

\section{Conclusions}
In summary, the effort to increase and optimizing the efficiency of heterojunction OPVs
necessarily involve many structural and energetics system parameters.
The above considerations focus on what is arguably the most important generic issue -
the optimization of the interfacial LUMO-LUMO gap that compensate between the need to
overcome the exciton binding energy and the required minimization of losses. For the present
simplified model and our choice of parameters we find the most efficient setup for interface
gap energies somewhat above $0.4\,\rm eV$. More important is the fact that the present model
with future generalizations that should include transport in the donor and acceptor phases
and polaronic relaxation following redistribution of charge densities provides a framework
for analyzing such efficiency measures. The mean field approach introduced here provides
a reliable approximation that can be used for fast evaluation of more complex model systems
and will be useful in extending these studies to realistic OPV cell models.
A more ambitious task would be to generalize the concept advanced in the article to the quantum regime. 
Consideration of quantum effects coherence (see, for example, Ref.~\onlinecite{Suclly:2010}) may prove useful in the discussion
of fundamental limits to photovoltaic efficiency.
\begin{acknowledgements}
  We thank Prof. Mark Ratner for helpful discussions. M.E. gratefully acknowledges funding by a
  Forschungsstipendium by the Deutsche Forschungsgemeinschaft (DFG, grant number EI 859/1-1).
  The research of A.N. is supported by the Israel Science Foundation, the Israel-US Binational Science
  Foundation, the European Science Council (FP7 /ERC grant no. 226628) and
  the Israel - Niedersachsen Research Fund.
\end{acknowledgements}


\begin{thebibliography}{99}
\bibitem{note}
    We take air mass 1.5 at standard test conditions (corresponds to $1000\,\rm
    {W}{m}^{-2}$) and assuming an illuminated area of $1\,\rm \mu m^2$.
\end{thebibliography}
\end{document}